# REDUCING PACKET OVERHEAD IN MOBILE IPv6

Hooshiar Zolfagharnasab[1]

[1]Department of Computer Engineering, University of Isfahan, Isfahan, Iran
hoppico@eng.ui.ac.ir
hozo19@gmail.com

## ABSTRACT

*Common Mobile IPv6 mechanisms, Bidirectional tunneling and Route optimization, show inefficient packet overhead when both nodes are mobile. Researchers have proposed methods to reduce packet overhead regarding to maintain compatible with standard mechanisms. In this paper, three mechanisms in Mobile IPv6 are discussed to show their efficiency and performance. Following discussion, a new mechanism called Improved Tunneling-based Route Optimization is proposed and due to performance analysis, it is shown that proposed mechanism has less overhead comparing to common mechanisms. Analytical results indicate that Improved Tunneling-based Route Optimization transmits more payloads due to send packets with less overhead.*

## 1. INTRODUCTION

Mobile IP is a technique enables nodes to maintain stay connected while they are moving through networks [1]. Due to Mobile IP protocol, a communication can be established between a Mobile Node (MN) and a Corresponding Node (CN) regardless to their locations.

The Mobile IP protocol supports transparency above the network layer including transport layer which consists of the maintenance of active TCP connections and UDP port bindings, and application layer. Mobile IP is most often found in wireless WAN environments where users need to carry their mobile devices across multiple LANs with different IP addresses [2]-[4]. Mobile IP is implemented in IPv6 via two mechanisms called Bidirectional tunneling and route optimization [1] [8].

In order to enable mobility over IP protocols, network layer of mobile devices should send messages to inform other devices about location and network they are wandering. Original packets from the network upper layers are embedded in packets containing mobile routing headers. Reducing mobility overhead causes more data to be sent with each packet. Therefore some mechanisms are used to reduce mobility overhead. In this paper, a new mechanism is proposed to reduce mobility overhead by reusing address field of IP address twice.

## 2. RELATED WORKS

Some attempts have been performed to improve security and performance in Mobile IP. C. Perkins proposed a security mechanism in binding updates between CN and MN in [5]. C. Vogt et al. in [6] proposed a proactive address testing in route optimization.

In other aspect, D. Le and J. Chang suggested reducing bandwidth usage due to use tunnel header instead of route optimization header when both MN and CN are mobile nodes [7].

It should be noted few papers focused on bandwidth reduction in Mobile IP while a lot of suggestions are proposed to solve issues in security and delay. In this paper, we are going to present a new technique to reduce bandwidth by diminishing overhead of packets when both MN and CN are mobile nodes.

## 3. MOBILE IPv6

Discussing about Bidirectional and route optimization (standard mechanisms), we will talk about their advantages and disadvantages. Later a method in [7] is presented to cover some disadvantages in standard mechanisms. The evaluation of discussed method called Tunneling-based Route Optimization is followed to show the improvement in bandwidth usage.

### 3.1. Bi-directional Tunneling

Based on idea of easily implementation of indirect mode in Mobile IPv4 [8], Bidirectional tunneling is presented in Mobile IPv6. In Bidirectional Tunneling, MN and HA are connected to each other via a tunnel, so signaling is required to construct a tunnel between MN and CN. Packets sent from CN to MN passes through HA before deliverance to MN. Intercepting all packets destined to MN, HA detects by Proxy Neighbor Discovery [9]. Since MN is not present in home network and assuming noticed tunnel is constructed, HA encapsulates each detected packet in a new packet addressed to MN's new care-of address (CoA) and sends them through the tunnel [10]. At the end of the tunnel, the tunneled packet is de-capsulated by MN's network layer before being surrendered to MN's upper layers.

Similar encapsulation is performed when MN sends packets. Encapsulated packets are tunneled to HA, that is called reverse tunneling, by adding 40 bytes as tunnel header, addressed from MN's CoA to HA. Being de-capsulated by HA, tunneling header is removed and modified packet is sent to CN through the Internet.

### 3.2. Route Optimization

In Route Optimization mechanism, packets are transmitted between MN and CN directly [3]. Binding Update (BU) messages are sent not only to HA, but also to all connected CNs to bind MN's current address to its HoA. Each CN has a table called Binding Cache to keep track of all corresponding MNs' CoA and their HoA. Similar table is kept in MN to determine whether a CN uses Bidirectional tunneling or route optimization. Also it is important to update CNs' binding cache by sending BU messages frequently.

Route Optimization mechanism uses Home Address Option header extension to carry MN's HoA when a packet is sent from MN to CN. Reversely when a packet is sent from CN to MN, another header extension called Type 2 Routing header is used.

Route optimization reduces the delay mentioned in Bidirectional tunneling. Putting routing task on each node, HA can handle more mobile nodes compared to Bidirectional tunneling.

In a scenario that both MN and CN are mobile nodes, route optimization can be implemented, too [1]. Since both MN and CN have HoA and CoA, packet routing requires both extension headers to carry enough information for the pair's network layer. Therefore, to transmit a packet from MN to CN, not only Home Address Option header, but also Type 2 Routing header should be filled with appropriate addresses. Since each extension header is 24 bytes, total overhead to transmit a packet between two mobile nodes is 48 bytes.

### 3.2. Tunneling-based Route Optimization

As discussed before, in a scenario when both MN and CN are mobile nodes, total overhead to carry a packet between nodes is 48 bytes in route optimization. To reduce the overhead, D. le and J. Chang in [7] proposed a mechanism called Tunneling-based Route Optimization (TRO). Like standard route optimization, TRO construct a tunnel to transfer packets directly between MN and CN. But in their proposed method, a Tunnel Manager is controlling packets. Not only tunnel manager is in touch with binding cache, but also it manipulates packets importing and exporting from the network layer.

As long as MN's transport layer create a packet from MN's HoA destined to CN's HoA, the packet is surrendered to MN's tunnel manager before it is sent. Since tunnel manager is aware of CN's mobility, it encapsulates the packet in a new packet addressed from MN's CoA to CN's CoA. Later the packet is sent through the tunnel to CN. At the other side of tunnel, CN's tunnel manger de-capsulate the packet, extracting the original packet addressed from MN's HoA to CN's HoA. Then the packet is surrendered to transport layer which is still unaware of mobility.

To maintain compatible with previous mechanisms, BU messages are changed. By using a flag called ROT, tunnel manager decides whether to use Tunneling-based Route Optimization or standard route optimization [7].

TRO mechanism benefits from using 40 bytes tunnel header instead of using 48 bytes extension header when standard route optimization is used. Result presented in [7] shows that TRO can increase performance in Mobile IP comparing to standard mechanisms.

## 4. IMPROVED TUNNELING-BASED ROUTE OPTIMIZATION

More reduction can be accessed in order to spend less header overhead in communication between MN and CN, when they are both mobile nodes. Each node constructs a binding cache to keep the address of the other, so there is no necessity to send HoA of the other pair via header extension because it can be obtained from binding cache by the help of CoA included in packet. In other words, header overhead is reduced by using IPv6 address fields twice, both for the Internet addressing and mobile addressing. Instead, a tunnel manager should be embedded not only to control binding cache, but also change the packet header. The tunnel manager should control whether IPv6 address header is used for Internet addressing or mobile addressing. Later in this section, we discuss about Improved Tunneling-based Route Optimization method.

### 4.1. Protocol Model in End-Points

Mobile IPv6 protocol should change a little to support overhead reduction. Both nodes should be devised with a tunnel manager which control and change all packets switched between MN and CN. Also the noticed tunnel manger should be allowed to access binding cache in order to find corresponding HoA of a node. Fig.1 depicts the protocol model in sender and receiver.

### 4.2. Improved Tunneling-based Route Optimization Routing

Below, we discuss two scenarios to explain our proposed method. It should be mentioned that a tunnel between MN and CN should be initiated at first. Also BU messages have been sent to construct binding cache in both CoA and HoA of the other pair. In a situation when CN is unaware of MN's new location, same action done in route optimization, is performed.

As long as MN wants to send a packet to CN, since mobility is transparent to upper layers in nodes, MN's network layer sets both source of the packet to MN' HoA and destination to CN's HoA. In the next step, when tunnel manager gets the packet, it updates the packet by changing both packet's source and destination. Since MN is in a foreign network, it changes the source field from its HoA to its CoA. Later, searching binding cache (by the help of CN's HoA), it finds CN's corresponding CoA and then writes it in the destination address field. Altered packet is sent directly to CN through the tunnel.

By reception of packet to the other side of the tunnel, CN's tunnel manager manipulates the packet to make it ready for upper layers. First manipulation is performed by changing the packet's destination from CN' CoA to CN's HoA. Next step is followed by searching binding cache with MN's CoA to find corresponding HoA. Later, the CN's tunnel manger then change packet's source from MN's CoA to what has just been found, MN's HoA. As long as changes are finished, the updated packet is surrendered to upper layers. Due to Fig. 1, packets sent from MN to CN are addressed as shown in Fig. 2.

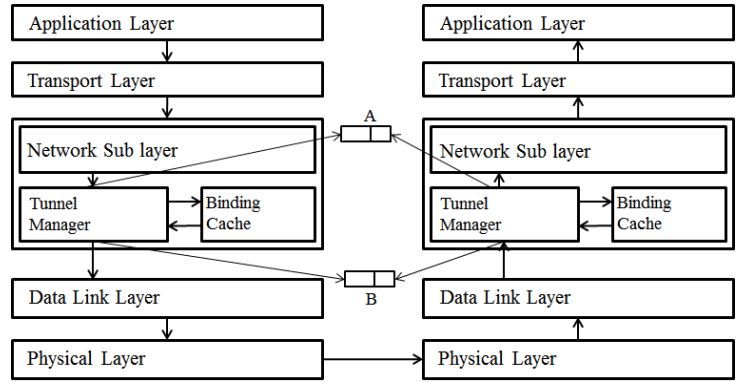

Figure 1. Protocol model for route optimization and packets passing between layers

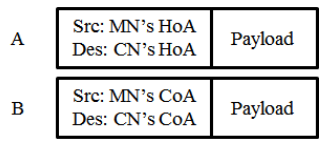

Figure 2. Improved tunneling-based route optimization packets due to Fig.1

Same action is performed when a packet is sent from CN to MN. Since CN's network upper layers are unaware of mobility, a packet is constructed which is addressed from CN's HoA to MN's HoA. As the packet is passed to CN's tunnel manger, due to binding cache, the destination of the packet is changed from MN's HoA to MN's CoA. Since CN knows its CoA, tunnel manger updates the packet's source from its HoA to CoA. Then the packet is tuneled to MN.

Similarly, MN's tunnel manager changes the packet's destination from MN's CoA to MN's HoA. Later, searching binding cache, the packet's source is also changed from CN's CoA to CN's HoA.

## 4.3. Changing BU messages

To maintain compatible with other MIPv6 mechanisms, binding messages should change. We propose to use two flags in order to distinguish three different mechanisms. Calling ROT0 and ROT1, these flags indicate whether route optimization or Tunneling-based Route Optimization or improved tunneling-based route optimization is used. Routing mechanisms due to ROT0 and ROT1 are listed in table 1.

Table 1. Routing mechanism due to ROT flags.

| Mechanism | ROT1 | ROT0 |
|---|---|---|
| Route Optimization | 0 | 0 |
| Tunneling-based Route Optimization | 0 | 1 |
| Improved Tunneling-based Route Optimization (proposed method) | 1 | 1 or 0 |

## 5. EVALUATION

Comparing to three other mechanisms, we evaluate our proposed method. Since Improved Tunneling-based Route Optimization mechanism intends to reduce header overhead, main comparison metric is bytes consumed to establish mobile communication. We used relation 1 proposed in [7] to calculate mobility overhead. It should be noted that mobility overhead is bytes used to establish mobility communication, and is different from overhead used to route packets through network layer.

$$Mobility\_overhead\_ratio = \frac{Mobility\_Addition\_Size}{Original\_Packet\_Size} \qquad (1)$$

Also as exception, comparing to Bidirectional tunneling mechanism, communicating time is also mentioned which is defined as total time for a packet to deliver from source to destination.

Moreover, packets are assumed to be 1500 bytes that is maximum transmission unit size in Ethernet, containing IPv6 packets, extension header if needed and tunneling overhead.

## 5.1. Comparing to Bidirectional tunneling

As mentioned before, in Bidirectional tunneling, packets from CN should be tunneled from HA to MN and are replied in the same tunnel from MN to HA, called reverse tunneling. For each time a packet is tunneled, 40 bytes are used to route the packet to the other side of tunnel. As a packet is tunneled twice to reach to destination, 80 bytes are consumed in two different communications. Total bandwidth which is used to carry a packet from source to destination is calculated as follows:

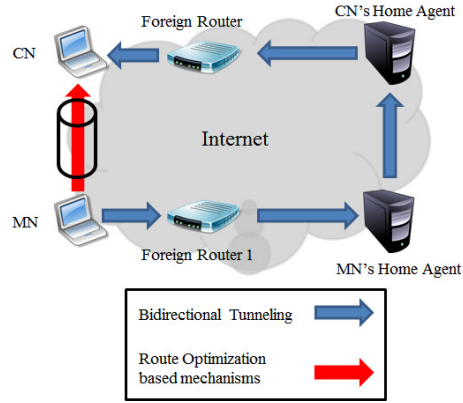

Figure 3. Comparing delay time for Bidirectional tunneling mechanism and route optimization based mechanisms

$$Mobility\_Overhead\_Ratio = \frac{Tunnel\_Header\_Size_{HA \to MN}}{Original\_Packet\_Size} + \frac{Tunnel\_Header\_Size_{MN \to HA}}{Original\_Packet\_Size}$$

$$= \frac{40}{1500 - 40} + \frac{40}{1500 - 40} = 5.48\% \qquad (2)$$

Also in Bidirectional tunneling, each routing elapses one Internet routing time [11] because each node can be anywhere in the Internet. Due to Fig.3, total delay consists of three Internet routing time that is calculated from:

$$Total\_Time = T_{MN \to HA_{MN}} + T_{HA_{MN} \to HA_{CN}} + T_{HA_{CN} \to CN} \cong 3 \times T_{Internet} \qquad (3)$$

In improved tunneling-based route optimization, since nodes are connected to each other through a tunnel, there is no obligation to tunnel packets twice between MN and HA. Also address field of packet is used both for tunnel and IPv6 header. Therefore, reduction in both overhead and delay are sensible. Mobility Overhead Ratio is calculated as follows:

$$Mobility\_Overhead\_Ratio = \frac{0\,B_{IPv6\_tunnel\_header}}{1500 - 0} = \frac{0}{1500} = 0\%$$ (4)

Also delay in proposed mechanism is computed from:

$$Total\_Time = T_{MN \to CN} \cong T_{Internet}$$ (5)

It means in Improved Tunneling-based Route Optimization mechanism s more efficient both in overhead and delay.

## 5.2. Comparing to Route Optimization

Although both route optimization and proposed mechanisms construct a tunnel to reduce delay and overhead regarded to communicate two mobile nodes, different overheads are used to route a packet in the constructed tunnel. In the situation when both nodes are mobile, route optimization uses Home Address Option and Type 2 routing extension headers as it is shown in Fig. 4. Since each extension header is 24 bytes in size, total mobility header added to IPv6 packet is 48 bytes. So mobility overhead ratio is calculated as follows:

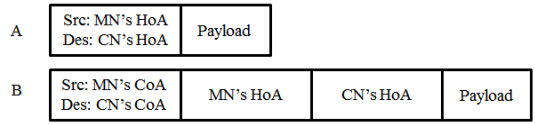

Figure 4. Route optimization packets due to Fig.1

$$Mobility\_Overhead\_Ratio = \frac{24\,B_{Type2} + 24\,B_{HoA\_Option}}{1500 - 48} = \frac{48}{1452} = 3.3\%$$ (6)

Since packets are tunneled directly to each node, one Internet time is needed for this communication due to Eq. 7.

$$Total\_Time = T_{MN \to CN} \cong T_{Internet}$$ (7)

Because Improved Tunneling-based Route Optimization uses address field of packet both for tunneling and IPv6 routing, as it calculated before, it uses 0% of total packet size.

Using same tunnel, total delay time is same for both route optimization and proposed method.

## 5.3. Comparing to Tunneling-based Route Optimization

Tunneling-based Route Optimization is proposed not only to decrease communication delay, but also to reduce overhead. It benefits from both tunneling idea used in Bidirectional tunneling and connecting directly used in route optimization. Tunneling header which is 40 bytes is added to IPv6 packet duo to reduce 48 bytes of extension headers. Fig. 5 shows packets A and B due to Fig. 1 when Tunneling-based Route Optimization mechanism is used. Also, mobility overhead ratio is calculated as follows:

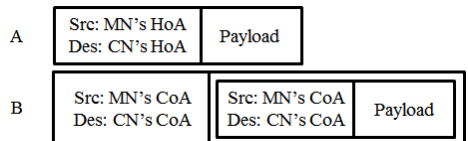

Figure 5. Tunneling-based Route Optimization packets due to Fig.1

$$Mobility\_Overhead\_Ratio = \frac{40\,B_{IPv6\_tunnel\_header}}{1500-40} = \frac{40}{1460} = 2.74\% \qquad (8)$$

Same total delay is calculated in Tunneling-based Route Optimization, because same tunnel is used to carry packets.

Comparing to Improved Tunneling-based Route Optimization mechanism, proposed method has no overhead in header used in mobile communication. And total delay is the same to route optimization mechanism.

Listed in table 2, Mobile IPv6 mechanisms are compared to each other. All in all it is obvious that proposed method can reduce both delay and bandwidth used in mobile nodes' communication.

Table 2. Comparison between Mobile IPv6 mechanisms

| Mechanism | Packet Overhead (%) | Delay (Internet Time) |
|---|---|---|
| Bidirectional Tunneling | 6.6 | 3 |
| Route Optimization | 3.3 | 1 |
| Tunneling-based Route Optimization | 2.74 | 1 |
| Improved Tunneling-based Route Optimization (proposed method) | **0** | **1** |

## 6. Conclusion

In this paper, performance of both standard Mobile IPv6 routing mechanisms and Tunneling-based Route Optimization are analysed. To reduce packet overhead, we proposed improved Tunneling-based Route Optimization mechanism. In order to maintain compatible with standard mechanisms, not only the tunnel manager should be changed, but also Binding Update messages must be altered. Comparing to Bidirectional tunneling, route optimization and tunneling-based route optimization shows that the packet overhead of proposed mechanism is reduced significantly. Therefore regarding to less overhead for each packet, more data can be transmitted through network via a Mobile IP communication.

**Authors**

HooshiarZolfagharnasab was born in Mar. 10[th] 1987. Graduated from high school in 2005, he entered to university in his beloved field, Computer Engineering. Four years later, he was graduated from B.Sc. in Computer Engineering, Hardware engineering and soon he was admitted in University of Isfahan to continue his career. In Nov. 15th, he was graduated from University of Isfahan as soon as he received his M.Sc. in Computer Engineering, Architecture of Digital Systems.

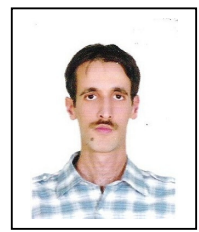

Now he is trying to find a PhD position to continue his career. He does all his effort to make the world a better place for all human beings.